\documentclass[aip,jcp,reprint,showpacs,amssymb,numerical]{revtex4-1}

\usepackage[english]{babel}
\usepackage{graphicx} %um Bilder einzufgen
\usepackage{amsmath} %fr bestimmte Formelumgebungen z.B. align
\usepackage{dsfont} %fr \mathds{1} (Einheitsmatrix - 1)
\usepackage[colorlinks=true,allcolors=black]{hyperref}
\usepackage{tikz}
\usetikzlibrary{calc}

\newcommand{\bra}[1]{\ensuremath{\left\langle #1\right|}}
\newcommand{\ket}[1]{\ensuremath{\left|#1\right\rangle}}

\newcommand{\tikzmark}[1]{\tikz[overlay,remember picture] \node (#1) {};}
\newcommand{\DrawBox}[1][]{
    \tikz[overlay,remember picture]{
    \draw[red,#1]
      ($(left)+(-0.2em,0.9em)$) rectangle
      ($(right)+(0.2em,-0.3em)$);}
}

\newcommand{\red}{\textcolor{red}}
\newcommand{\blue}{\textcolor{blue}}

\begin{document}
\graphicspath{{pictures/}}
\title{Bloch-Redfield equations for modeling light-harvesting complexes}

\author{Jan Jeske}
\affiliation{Chemical and Quantum Physics, School of Applied Sciences, RMIT University, Melbourne, 3001, Australia}

\author{David J. Ing}
\affiliation{Chemical and Quantum Physics, School of Applied Sciences, RMIT University, Melbourne, 3001, Australia}

\author{Martin B. Plenio}
\affiliation{Institut f\"ur Theoretische Physik, Albert-Einstein-Allee 11, Universit\"at Ulm, D-89069 Ulm, Germany}

\author{Susana F. Huelga}
\affiliation{Institut f\"ur Theoretische Physik, Albert-Einstein-Allee 11, Universit\"at Ulm, D-89069 Ulm, Germany}

\author{Jared H. Cole}
\affiliation{Chemical and Quantum Physics, School of Applied Sciences, RMIT University, Melbourne, 3001, Australia}

\begin{abstract}
We challenge the misconception that Bloch-Redfield equations are a less powerful tool than phenomenological Lindblad equations for modeling exciton transport in photosynthetic complexes. This view predominantly originates from an indiscriminate use of the secular approximation. We provide a detailed description of how to model both coherent oscillations and several types of noise, giving explicit examples. All issues with non-positivity are overcome by a consistent straightforward physical noise model. Herein also lies the strength of the Bloch-Redfield approach because it facilitates the analysis of noise-effects by linking them back to physical parameters of the noise environment. This includes temporal and spatial correlations and the strength and type of interaction between the noise and the system of interest. Finally we analyze a prototypical dimer system as well as a 7-site Fenna-Matthews-Olson (FMO) complex in regards to spatial correlation length of the noise, noise strength, temperature and their connection to the transfer time and transfer probability.
\end{abstract}

\maketitle
 
\section{Introduction}
Following experimental evidence of long-lived oscillatory features in the dynamical response of several photosynthetic systems \cite{Engel2007, Lee2007, Collini2010}, there have been many studies analyzing quantum coherence and the effects of decoherence in light harvesting complexes (LHCs) \cite{Huelga2013}. A complete understanding of the coherence properties and the spectral response of this type of molecular aggregates has been shown to require the inclusion of non-Markovian effects \cite{Jang2008, Ishizaki2008, Rebentrost2009nonmarkovian, Thorwart2009, Ritschel2011, Kolli2011} and/or the use of numerically exact methods \cite{Ishizaki2009b, Kreisbeck2012, Prior2010, NalbachArxiv}. However, Markovian master equations, such as Lindblad or Bloch-Redfield equations, still provide powerful insight and generally serve as the initial test-bed from which more accurate descriptions can be built up \cite{Plenio2008, Mohseni2008, Rebentrost2009, Caruso2009, Fassioli2010, Hoyer2010, Sinayskiy2012}. 

While the Bloch-Redfield approach has been successfully used by several groups\cite{Renger1998, Renger2002, Novoderezhkin2004, Kjellberg2006, Adolphs2006, Abramavicius2011, Rebentrost2009SpatialCor} to describe excitonic dynamics in LHCs, there have also been many publications emphasizing its shortcomings\cite{Palmieri2009, Ishizaki2009, Caruso2009, Rebentrost2009nonmarkovian, Nalbach2010, Chen2010, Huo2010, VanVeenendaal2010, Huo2011a, Sarovar2011, Renaud2012, Shabani2012}. In this paper, we show how to apply Bloch-Redfield theory, such that (i) an underlying physical model can guarantee physically acceptable time evolution, (ii) the secular approximation does not have to be applied, (iii) if it is applied carefully it preserves the couplings between populations and coherences which lead to coherent oscillations between sites and (iv) the Bloch-Redfield equations are not less general than Lindblad equations and offer a greater variety of noise models than phenomenological Lindblad equations.

Lindblad equations guarantee complete positivity of the corresponding dynamical map, i.e.~a physical time evolution, by their mathematical form. The generator of the time evolution corresponds to a quantum dynamical semigroup \cite{Lindblad1976, Gorini1976, Breuerbook, Rivasbook}, as compared to the group property obeyed by unitary time evolution of closed systems. This enables one to get physical results for an open quantum system without the necessity of deriving a master equation from a microscopic model. While phenomenological Lindblad equations are the mathematically simplest tool, an alternative is the Bloch-Redfield approach~\cite{Bloch1957, Redfield1957, Soares2011, Jeske2013formalism} which allows more comprehensive modeling options, such as the inclusion of noise correlations, besides rendering the correct thermal state at long times. The Bloch-Redfield equations describe an approximately Markovian dynamics. Strong non-Markovian effects, which can for example arise from coupling to phonons with large reorganisation energies\cite{Ishizaki2009c}, can only be taken into account by including the causes of such effects (for example such phonons) into the system Hamiltonian and describing the combined system by Markovian dynamics\cite{Plenio2013}.

The Bloch-Redfield formalism requires a closer connection to the microscopic parameters and the causes of decoherence. Within the Bloch-Redfield approach, the evolution of the reduced system density matrix is determined by the explicit form of the system-environment interaction Hamiltonian and the environment's spatial and temporal correlations.

Simplified versions of the Bloch-Redfield equations (sometimes labelled `Redfield equations') have previously been applied in a way that caused issues such as non-positivity or divergences in the time evolution, which were solved by an application of the overly indiscriminate full secular approximation. This in turn caused the inability to model coherent oscillations as a result of decoupling the evolution of exciton populations and coherences. Here, populations refer to the diagonal elements $\rho_{jj}$ of the density matrix $\rho$ and coherences to the off-diagonal elements $\rho_{jk}$. This decoupling of populations and coherences has been misinterpreted as an inherent problem of Bloch-Redfield equations in the secular approximation\cite{Kjellberg2006, Palmieri2009, Rebentrost2009nonmarkovian, Caruso2009, Chen2010, Huo2010}. These issues have lead to the misconception that Bloch-Redfield equations are insufficient to model relevant aspects of the dynamics of light-harvesting systems or are a less general case than Lindblad equations\cite{Palmieri2009, Nalbach2010, Sarovar2011, Renaud2012}. This view is propagating through the literature, leading some authors to discount the Redfield equations unjustly.  Statements such as ``the full Redfield theory ... usually results in unphysical density matrices with negative or diverging populations'' \cite{Palmieri2009}, ``... yields an unphysical density matrix: populations may ... diverge"\cite{Abramavicius2010}, and "... the Redfield ... equations, which can in some situations, provide inaccurate descriptions of these processes due to various approximations employed (e.g., use of ... the secular approximation)" \cite{Huo2011a} show how the Redfield equations are discounted due to this misunderstanding.

Here we present a general method to model the time evolution and transport dynamics of LHCs with Bloch-Redfield equations which has none of these pathologies and enables comprehensive modeling options for environmental parameters. Furthermore we state explicitly the rules of the secular approximation which ensure that coherent oscillations are not lost. As a simple example, we apply our formalism to a model dimer system and show how coherent oscillations and decay arise from the system and noise environment parameters. 

One important advantage of the Bloch-Redfield approach is that both temporal and spatial correlations in the environment can be modelled, assuming that the effect of the environment is approximately Markovian. Using other models, the relevance of spatial correlations was pointed out in \cite{Nalbach2010, Fassioli2010, Pelzer2013}, as they were found to enhance coherence\cite{Rebentrost2009SpatialCor, Sarovar2011, Abramavicius2011}, which is consistent with the more general result that a decoherence-free subspace emerges in the single-excitation subspace for strongly correlated environments\cite{Jeske2013formalism, Jeske2013spinchain, Jeske2014metrology}. Some models found noise correlations to slow transport down \cite{Sarovar2011}, others to speed up transport up to a certain optimal value depending on the strength of the reorganisation energy\cite{Wu2010}. The latter is in agreement with our findings in this paper using Bloch-Redfield equations. We show that the Fenna-Matthews-Olson (FMO) complex achieves maximum efficiency for a parameter regime with finite spatial noise correlations.

\section{Lindblad vs. Bloch-Redfield equations}
In light-harvesting complexes the system Hamiltonian $H_s$ describes the system and all its quantum coherent features, typically the electronic degrees of freedom. In terms of the states $\ket{j}$ which describe an excitation on the spatial site $j$, the system Hamiltonian is typically of the form $H_s=\sum_j \epsilon_j \ket{j}\bra{j} + \sum_{j\neq k} g_{jk} (\ket{j}\bra{k}+\ket{k}\bra{j}$, where the first term describes the sites at energy $\epsilon_j$ and the second term the dipolar couplings between sites $g_{jk}$. To then model the noise influence and incoherent dynamics due to electron-phonon coupling resulting from the phonon environment of intramolecular and protein vibrations a master equation approach is often used. The Lindblad equations are a popular tool for this because their mathematical form guarantees a time evolution that is linear trace preserving and positive (completely positive, to be precise, as we will discuss in section \ref{sec mapping to lindblad}), restricting the populations, i.e.~the diagonal density matrix elements, to be positive and always sum to one. This is physically necessary since they represent the probabilities of measuring the corresponding state. However modeling beyond Lindblad equations opens up new capabilities and does not need to violate positivity \cite{Whitney2008}.

\begin{figure}
\flushleft
%\vspace{1.6cm} %\\
\includegraphics{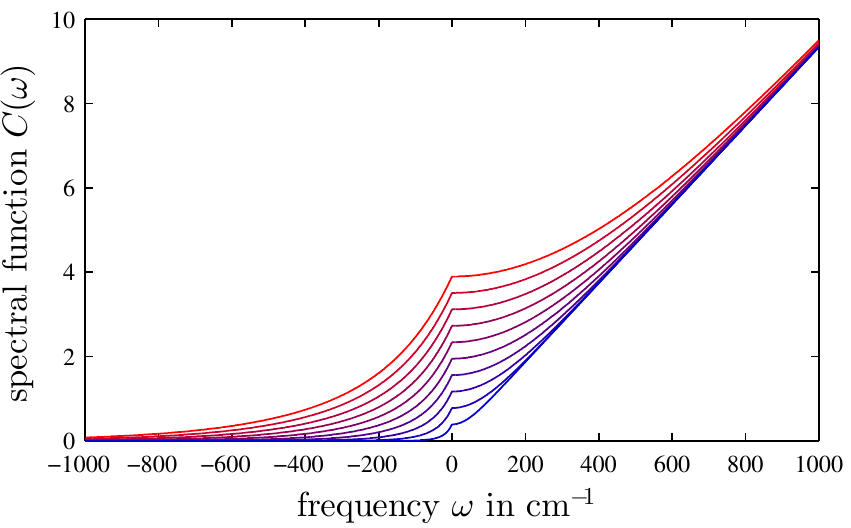}
\caption{Ohmic spectral function for different temperatures $T$ between 30 K (blue) and 300 K (red). The noise around zero frequency determines the dephasing strength and is strongly temperature dependent. The noise at higher frequencies determines the recombination and is largely unaffected by temperature. For negative frequencies the spectral function is exponentially damped which leads to condition for the detailed balance being preserved.}
\label{fig spectral}
\end{figure}

In difference to the Lindblad equations, the mathematical form of Bloch-Redfield equations does not guarantee \emph{a priori} positivity of the density matrix, i.e.~physicality. It is an underpinning consistent microscopic noise model that guarantees physical behaviour. This is however where the strength of the Bloch-Redfield formalism lies. It connects system behaviours to physical properties of the noise and the system-noise interaction type. Specifically the equations derive from an interaction Hamiltonian 
\begin{align}
H_{int}=\sum_j s_j \otimes B_j, \label{interaction Hamiltonian}
\end{align}
where $s_j$ are system operators and $B_j$ bath operators and $j$ runs over the spatial sites in the system. The spectral function $C_{jk}(\omega)$ of the noise environment defines both the noise spectrum and the strength of spatial correlations between sites $j$ and $k$. The spectral function is given by the Fourier transform of the bath correlation function and is different from the spectral density of an environment, which defines the number and coupling strength of environmental fluctuators at a certain frequency and is temperature-independent. The spectral function on the other hand is dependent on temperature and frequency as illustrated in figure~\ref{fig spectral}. To model a Markovian environment, the spectral function must be slowly changing with frequency on the scale set by the system evolution.  This is a key point in correctly deriving physical master equations, which will be detailed further in section~\ref{sec detailed balance}.

While the Bloch-Redfield equations can be written down in any basis of states, one still needs to find the eigenbasis of the system Hamiltonian $V^\dagger H_S V = \text{diag}(\varepsilon_1, \varepsilon_2, \dots)$. The general form of the Bloch-Redfield equations is then given by\cite{Jeske2013formalism}:
\begin{multline}
\dot \rho = \frac{i}{\hbar} [\rho,H_s] + \frac{1}{\hbar^2} \sum_{j,k} \left(-s_j V q_{jk} V^\dagger \rho + V q_{jk} V^\dagger \rho s_j 
\right.\\ \left. 
-\rho V \hat q_{jk} V^\dagger s_j + s_j \rho V \hat q_{jk} V^\dagger\right) \label{B-R equations}
\end{multline}
\begin{align}
&\text{with} \nonumber\\
&\langle a_n| q_{jk} |a_m\rangle = \bra{a_n} V^\dagger s_k V \ket{a_m} \frac{1}{2} C_{jk}(\omega_m-\omega_n) \label{q-elements}\\
&\langle a_n| \hat q_{jk} |a_m\rangle = \bra{a_n} V^\dagger s_k V \ket{a_m} \frac{1}{2} C_{kj}(\omega_n - \omega_m) \label{qhat-elements}
%\\ &C_{jk}(\omega)= \int_{-\infty}^\infty d \tau \; e^{i \omega \tau} \, \langle  \tilde B_j(\tau) \tilde B_k(0) \rangle \label{spectral function definition}
\end{align}
where $\ket{a_n}$ can be any basis of states of the quantum system; if the eigenstates of the system Hamiltonian $H_S$ are chosen, then the matrices $V$ become unity operators. The system operators $s_j$ define which part of the system couples to the noise environment. There can be noise on site-operators (e.g. site-energy-noise $\sigma_z^{(j)}=2\ket{j}\bra{j}-\mathds{1}$, site recombination $\sigma_x^{(j)}=\ket{j}\bra{0}+h.c.$) or there can be noise on the coupling operators between sites (e.g.~transversal couplings $\ket{j}\bra{k}+\ket{k}\bra{j}$, longitudinal couplings $\sigma_z^j \sigma_z^k$) \cite{Huo2012}. 

When the elements of the density matrix are rewritten as a column-ordered vector the Bloch-Redfield equations can be written as a matrix multiplication with this vector. This so-called superoperator form reads:
\begin{align}
\dot{\vec \rho}&=\mathfrak{R} \vec \rho \label{superoperator form}\\
&=\frac{i}{\hbar} \left( H^T \otimes \mathds{1} - \mathds{1} \otimes H \right) \vec \rho \\
&\quad + \frac{1}{\hbar^2} \sum_{jk} \left( -\mathds{1} \otimes s_j V q_{jk} V^\dagger + s_j^T \otimes V q_{jk} V^\dagger \right) \vec \rho \nonumber\\
&\quad - \frac{1}{\hbar^2} \sum_{jk} \left( s_j^T V^* \hat q_{jk}^T V^T \otimes \mathds{1} + V^* \hat q_{jk}^T V^T \otimes s_j \right) \vec \rho \nonumber
\end{align}
where $\mathds{1}$ is the unity matrix of the dimension of $H$ and $\mathfrak{R}$ is the so-called Redfield tensor in matrix form.

\subsection{Measurement basis and oscillations}
\label{sec measurement basis}
In light-harvesting complexes there are typically two different types of evolution that are being modelled: different spectroscopy experiments in the lab on the one hand and the excitation dynamics as it occurs in nature and facilitates transport towards the reaction centre on the other. In the current manuscript we focus on the latter, which is a dynamic process between the spatial sites of the FMO complex. Populations should therefore be taken in the site basis (aka bare basis, defined as the eigenbasis of the Hamiltonian without couplings between sites). Then the populations correspond to the probability of finding an excitation at the corresponding site. If on the other hand the populations are taken in the excitonic basis, i.e. the eigenbasis of the Hamiltonian\cite{Palmieri2009} (with couplings) their physical meaning is unclear and oscillations in this basis \emph{do not necessarily} correspond to excitation transfer any more. Secondly such oscillations are not caused by the system dynamics any more since it is the basis of stationary states in the coherent system dynamics. The eigenbasis of the full Hamiltonian has a more natural place in understanding the system's reaction to external pulse sequences of 2D spectroscopy.

\subsection{Secular approximation}
The Bloch-Redfield equations can be simplified by means of a secular approximation. This can be hugely advantageous in order to find analytical solutions to the equations; for purely numerical solutions however it is not necessary and usually not advantageous because it should only neglect elements which do not significantly alter the time evolution. An overly indiscriminate application of the secular approximation does change the time evolution and is sometimes used because it can rid the equations of any physical inconsistencies should they arise from a physically contradictory choice of operators, spectral function and/or correlations. However, this will also inevitably lead to a complete decoupling of coherences and populations, which leads to the loss of coherent oscillations that has been stated so often. The central point is that for numerical simulations the secular approximation is unnecessary when the Bloch-Redfield equations are supported with a consistent underlying physical model as offered by equations \ref{interaction Hamiltonian} to \ref{qhat-elements} because it automatically leads to a physical time evolution. 

The secular approximation is useful to solve equations analytically and can help to map the Bloch-Redfield equations to Lindblad form\cite{Jeske2013formalism, Breuerbook}. Next we will show how to apply the secular approximation carefully in order not to change the dynamics significantly and to preserve the populations-coherence coupling relevant to coherent oscillations. The secular approximation should always be based on the existence of different, separated (i.e. secular) scales of system energies and frequencies. Such different scales can make some small elements of the superoperator $\mathfrak{R}$ negligible. Which elements can be neglected is different for each particular system and the following careful pairwise comparison of the elements is necessary. If the magnitudes of two diagonal elements in the superoperator $\mathfrak{R}$ differ by much more than their shared off-diagonal elements then these off-diagonal elements can be replaced with zero. We can restate this mathematically as:
\begin{align*}
\text{If }|\mathfrak{R}_{jj}|-|\mathfrak{R}_{kk}| \gg |\mathfrak{R}_{jk}|,|\mathfrak{R}_{kj}| \text{, then set }\mathfrak{R}_{jk}=\mathfrak{R}_{kj}=0
\end{align*}
Only those elements of the superoperator are set to zero for which the condition holds. We refer to this as the ``partial secular approximation''.  It preserves populations-coherence coupling and allows for non-monotonic excitonic population dynamics and coherent oscillations in any basis. While some work in the literature follows this careful procedure, others employ the overly indiscriminate ``full secular approximation'', which sets all dependencies between coherences and populations to zero regardless of the condition above. The full secular approximation only leaves those off-diagonal superoperator elements non-zero, which connect different populations. In superoperator form (eq.~\ref{superoperator form}) the vector $\vec{\rho}$ contains both coherences and populations at different positions $\rho_j$. We can restate the full secular approximation as:
\begin{align*}
\text{If } \rho_j \text{ or } \rho_k \text{ is a coherence, then set }\mathfrak{R}_{jk}=\mathfrak{R}_{kj}=0
\end{align*}
We emphasize once again that this full secular approximation is usually overly indiscriminate as it sets non-negligible elements to zero, thereby changing the time evolution significantly. Although this extreme case is guaranteed to rid the equations of any non-physical inconsistencies, it usually also rids the system of dynamically important and physically well justified coherent oscillations. The frequent use of this full secular approximation has led to the claim that Redfield theory can not reproduce oscillations which are caused by dependency of populations and coherences\cite{Palmieri2009}. An example in section \ref{sec applying secular} will show how these physically relevant dependencies are preserved by the partial secular approximation.

The occurrence of two separated energy scales is quite common in prototypical light-harvesting systems. This is because the excitonic (on-site) energy, which is typically of the order 10,000 cm$^{-1}$, is much larger than the coupling energies and the differences of the on-site energies, which are both typically on the order of 1 cm$^{-1}$ to 100 cm$^{-1}$. A secular approximation based on this difference leads to two very general results: Firstly, it decouples the one-excitation subspace from both the ground state and from states with two or more excitations. This means that for modeling excitation transport in LHCs as it occurs in nature one can neglect states with more than one excitation in the time evolution. Multiple-excitation states then only need to be considered when modeling 2D spectroscopy with pulse sequences that create multiple excitons. Secondly, the secular approximation separates the bath of the longitudinal couplings (e.g.\ $s_j=\sigma_z^{(j)} = 2\ket{j}\bra{j}-\mathds{1}$) from the bath of the transversal couplings (e.g.\ $s_j=\sigma_x^{(j)} = \ket{j}\bra{0} + h.c.$) where $\ket{0}$ is the system's ground state with no excitation. This simplifies modeling since correlations in the noise of these different types of coupling do not have any effects and can be neglected.

\subsection{Detailed balance}
\label{sec detailed balance}
In the thermal equilibrium state, the populations of two sites are given by $\rho_{11}/\rho_{22}=e^{-\hbar \epsilon_{12}/k_B T}$, where $\epsilon_{12}$ is the energy difference of the two sites. This detailed balance condition translates in the Bloch-Redfield formalism to the property of the spectral function $C(-\omega)=e^{- \hbar \omega/k_B T} C(\omega)$, where $\omega>0$. Spectral functions derived from a microscopic model such as the spin-boson or related models have this property already \cite{Leggett1987, Zanardi1998, Jeske2013formalism}. Regarding the detailed balance condition at low temperatures, the spectral function for negative frequencies can be approximated by zero since for low temperatures $e^{- \hbar \omega/k_B T}\approx0$. This needs to be done in a manner consistent with the Markov approximation. The Markov approximation requires that the spectral function does not change on the scale relevant to the system dynamics (typically $g_{jk}$). Therefore one \emph{must} obey $C(0 \pm g_{jk})\approx const$. This is one of the subtle details which can cause issues with non-positivity. If there are other larger scales (e.g. $\omega_j \gg g_{kl}$) the spectral function has no restrictions and can vary appreciably on this scale.

The detailed balance is consistent with the fact that excitons recombine but are not spontaneously created even at room temperature from the surrounding noise environment. This means once recombination is considered the long-time equilibrium will have all population in the ground state even at room temperature. Neglecting recombination the detailed balance can be applied to the single excitation subspace only \cite{Palmieri2009}, however this is somewhat artificial since the detailed balance is typically driven by energy-exchanging noise on each site, i.e.~recombination noise.

\subsection{Spatial correlations}
The option to model spatially correlated noise arises naturally in the formalism through the spectral function $C_{jk}(\omega)$. It allows for a spatially decaying correlation function with a distinctive correlation length $\xi$, i.e.~one can transition smoothly between infinite, finite and no spatial correlations. For example, we can model exponentially decaying spatial correlations for a three-dimensional model as:
\begin{align}
C_{jk}(\omega)=\exp\left(-\frac{|\bf{r}_j - \bf{r}_k|}{\xi}\right) C(\omega)\label{spatial correlations exp decay}
\end{align}
Using this form of spatial correlations we can describe systems with several sites (e.g.~the realistic 7-site FMO model, see section \ref{sec effects in large LHCs}), where finite correlation length can be applied to the actual geometry of the LHCs.

\begin{figure}
\flushleft
%\vspace{1.4cm}                                %\\
\includegraphics{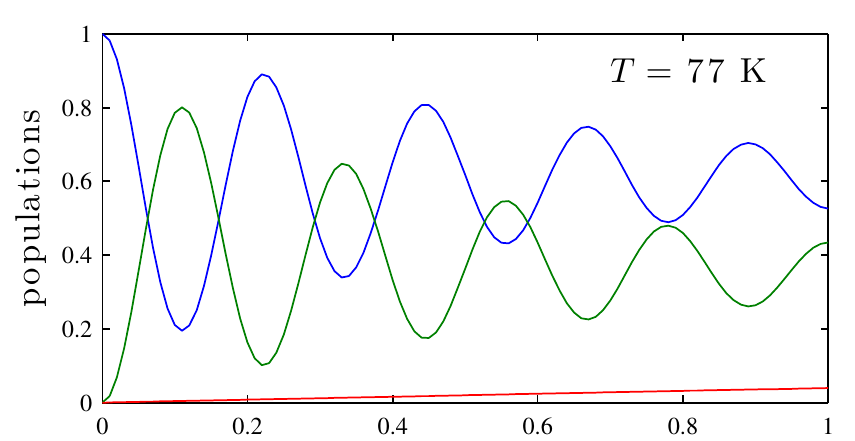}\\
\includegraphics{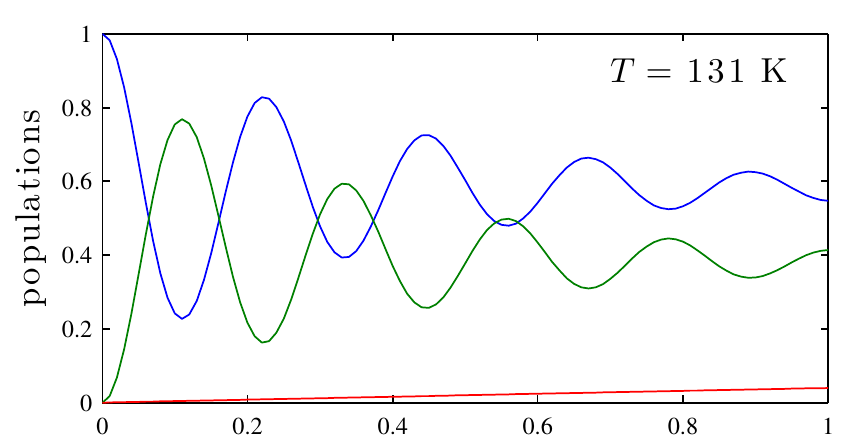}\\
\includegraphics{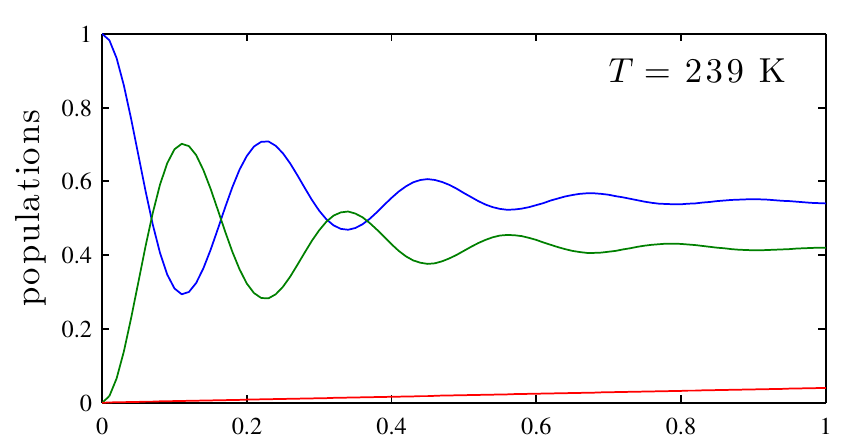}\\
\includegraphics{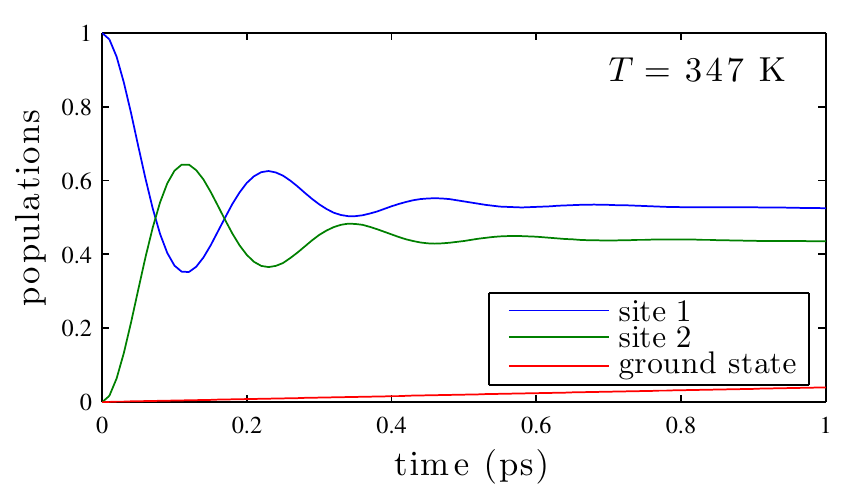}
\caption{Dynamics of a model dimer system at different temperatures $T$ using the Bloch-Redfield approach. The oscillations decay due to dephasing. This is stronger at higher temperatures. Recombination makes the populations in the two sites slowly decay and the population of the ground state (red) rise. System-environment couplings are $v=5 $cm$^{-1},  \nu=0.1 $cm$^{-1}$. The secular approximation was not needed to produce these plots.}
\label{fig dimer evolution}
\end{figure}

\section{Model example: a dimer system}
\label{sec dimer system}
As an illustrative example, we consider a dimer system with Hamiltonian 
\begin{align}
H_s=\left( \begin{array}{cc} \epsilon_1& g \\ g & \epsilon_2 \end{array} \right) 
\end{align}
This simple system of two sites (with on-site energies $\epsilon_1, \epsilon_2$) serves as the most basic model in many photosynthetic systems\cite{Palmieri2009}, e.g.~in the reaction centre of the purple bacteria Rhodobacter sphaeroides \cite{Lee2007, Huo2012} or in water soluble chlorophyll proteins (WSCP) \cite{Theiss2007} and serves as the most basic model of excitation transfer in light-harvesting complexes.  

Neglecting all noise influences, an excitation will oscillate between the sites with frequency $\hbar \omega = \sqrt{g^2+(\epsilon_1-\epsilon_2)^2/4}$. The oscillation amplitude will be strongest for $\epsilon_1-\epsilon_2=0$. In other words, the stronger $g$, the faster the oscillations, but increasing $|\epsilon_1-\epsilon_2|$ decouples the sites. 
This role of off-diagonal couplings and diagonal on-site-energies in the system Hamiltonian generalizes to more complex multiple-site systems. 

\subsection{Constructing the superoperator}
Next we consider dephasing noise coupling to the site-energies by setting the system operators in the interaction Hamiltonian to $s_1= 2v\ket{1}\bra{1}-v\mathds{1}$, $s_2=2v\ket{2}\bra{2}-v\mathds{1}$, where $v$ is the coupling strength. First we model spatially uncorrelated noise and an Ohmic spectral function\cite{Leggett1987, Reina2002, Ishizaki2008, Rebentrost2009SpatialCor, Nalbach2010, Hoyer2010} $C_{jk}(\omega)=\alpha \omega \coth(\hbar \omega/2 k_B T)\, \delta_{jk} $, where $\alpha$ accounts for the noise strength (see figure \ref{fig spectral}). The oscillations then show an envelope exponential decay, due to the loss of phase coherence between the two sites. For sites with similar energy $|\epsilon_1-\epsilon_2| \ll g$ the decay rate is given by $\gamma_2=v^2 [C(2g)+C(-2g)]/2$.  Since $C(\omega)$ hardly changes on the scale of $g$ one can approximate $C(\pm 2g)\approx C(0)\propto T$ for the given spectral function. In figure \ref{fig dimer evolution} we can see a few numerical examples for different temperatures. With decreasing temperature the environmental fluctuations diminish, i.e.~dephasing noise on the system is reduced and coherent oscillations last for longer times. The oscillation frequency is not affected by temperature since we have not considered a temperature dependency of the system Hamiltonian $H_S$. 
%calculations in dimer.nb

We then add recombination noise into our considerations. To do so we need to add the ground state, in which the excitation has vanished from all sites to the system Hamiltonian. 
\begin{align}
H_s=\left( \begin{array}{ccc} \epsilon_1& g&0 \\ g & \epsilon_2 &0\\ 0&0&\epsilon_0\end{array} \right) 
=\left( \begin{array}{ccc} 0& 71.3&0 \\ 71.3 & 46.4 &0\\ 0&0&-12210\end{array} \right) 
\end{align}
The energy $\epsilon_0$ of the ground state $\ket{0}$ is one excitonic energy lower than $\epsilon_1$ and $\epsilon_2$. This difference is typically two orders of magnitude larger\cite{Hoyer2010} than all other relevant parameters in $H_S$. Therefore any couplings between the ground state and states $\ket{1}$ and $\ket{2}$ would only have a negligible effect and are neglected entirely by employing the rotating wave approximation/ secular approximation. For simplicity we set them to zero in the first place. We then extend the system operators of each site $s_1= 2v\ket{1}\bra{1}-v\mathds{1} + \nu \ket{1}\bra{0}+\nu \ket{0}\bra{1},$ $ s_2=2v\ket{2}\bra{2}-v\mathds{1} + \nu \ket{2}\bra{0}+\nu \ket{0}\bra{2}$ and use the same spectral function as before. The excitonic energy is then lost from the system at a rate of $\gamma_1=\nu^2 C(\epsilon_{H}-\epsilon_0)$, where the spectral function is again approximately constant around this frequency. While the dephasing rate is of about the same order as the inter-site couplings, the recombination rate is typically much slower and on the order of 1 ns \cite{Owens1987, Rebentrost2009SpatialCor}. The recombination can be seen by a slow decay of the populations in sites 1 and 2 to the ground state in figure \ref{fig dimer evolution}. As the ground state is considerably lower in energy, almost all population is found in the ground state at thermal equilibrium, which is reached for very long times. This corresponds to the fact that the creation and existence of an exciton itself is a non-equilibrium process. However the recombination processes are typically much slower than dephasing processes due to a weaker noise coupling $\nu<v$. Therefore the system dephases first and then decays to the ground state on a longer time scale. In contrast to dephasing the recombination strength is almost temperature independent. In an Ohmic noise environment it is purely proportional to the excitonic energy.

\subsection{Applying the secular approximation}
\label{sec applying secular}
The secular approximation is based on two largely different (i.e.~`secular') scales of parameters involved. Usually these two scales are the large excitonic energy on the one hand and the small couplings between sites and noise strength on the other. The approximation should never alter the solutions significantly but merely simplify the process of finding a solution by setting those elements of the superoperator to zero, which only have a negligible effect on the solution. This is true for those off-diagonal superoperator elements whose magnitude is much smaller than the difference of their corresponding diagonal elements\cite{Jeske2012}. 

We demonstrate the validity of this method numerically for the considered dimer system. To write the superoperator as a matrix we first reorder the density matrix as a vector. We do so and arrange the elements so that the diagonal density matrix elements come first in the vector (eq.~\ref{dimer superoperator}). The Bloch-Redfield equations are then given in eq.~\ref{dimer superoperator} without the secular approximation and in eq.~\ref{dimer superoperator with secular} with the secular approximation. To transform from one to the other one needs to compare each pair of diagonal elements and if their difference is of the order $10,000$ then the two corresponding off-diagonal elements are set to zero. This detailed procedure yields what we call the partial secular approximation. Note that some elements connecting populations and coherences are left non-zero in the partial secular approximation. These are the blue off-diagonal superoperator elements outside the red rectangle in eq.~\ref{dimer superoperator} and~\ref{dimer superoperator with secular}. The full secular approximation would set all off-diagonal elements which are outside the red rectangle to zero and only leave behind those off-diagonal elements which link the populations $\rho_{11},\rho_{22}$ and $\rho_{33}$ (i.e.~the red elements inside the red rectangle). 
This full secular approximation would significantly alter the coherent evolution of the system. It would neglect any coherent oscillations and only leave the transition rates un-altered. 

\begin{widetext}
\begin{align}
\left( \begin{array}{c} \dot \rho_{11} \\ \dot \rho_{22} \\ \dot \rho_{33} \\ \dot \rho_{12} \\ \dot \rho_{13} \\ \dot  \rho_{21} \\ \dot \rho_{23} \\ \dot \rho_{31} \\ \dot \rho_{32} \end{array} \right)
=\left(
\begin{array}{ccccccccc}
 \tikzmark{left}-4 & \red{0} & \red{4} & \blue{0-71 i} & 3 & \blue{0+71 i} & 0 & 3 & 0 \\
 \red{0} & -4 & \red{4} & \blue{0+71 i} & 0 & \blue{0-71 i} & 3 & 0 & 3 \\
 \red{4} & \red{4} & -8\tikzmark{right} & 0 & -3 & 0 & -3 & -3 & -3 \\
 \blue{-1-71 i} & \blue{1+71 i} & 0 & -8-46 i & -1 & 0 & 0 & 0 & -1 \\
 2 & 0 & -2 & -2 & -9+12210 i & 0 & \blue{0+71 i} & 4 & 0 \\
 \blue{-1+71 i} & \blue{1-71 i} & 0 & 0 & 0 & -8+46 i & -1 & -1 & 0 \\
 0 & 2 & -2 & 0 & \blue{0+71 i} & -2 & -9+12256 i & 0 & 4 \\
 2 & 0 & -2 & 0 & 4 & -2 & 0 & -9-12210 i & \blue{0-71 i} \\
 0 & 2 & -2 & -2 & 0 & 0 & 4 & \blue{0-71 i} & -9-12256 i
\end{array}
\right)
\left( \begin{array}{c} \rho_{11} \\ \rho_{22} \\ \rho_{33} \\ \rho_{12} \\ \rho_{13} \\ \rho_{21} \\ \rho_{23} \\ \rho_{31} \\ \rho_{32} \end{array} \right)
\label{dimer superoperator}
\DrawBox[thick]
\\
\left( \begin{array}{c} \dot \rho_{11} \\ \dot \rho_{22} \\ \dot \rho_{33} \\ \dot \rho_{12} \\ \dot \rho_{13} \\ \dot  \rho_{21} \\ \dot \rho_{23} \\ \dot \rho_{31} \\ \dot \rho_{32} \end{array} \right)
=\left(
\begin{array}{ccccccccc}
 \tikzmark{left}-4 & \red{0} & \red{4} & \blue{0-71 i} & 0 & \blue{0+71 i} & 0 & 0 & 0 \\
 \red{0} & -4 & \red{4} & \blue{0+71 i} & 0 & \blue{0-71 i} & 0 & 0 & 0 \\
 \red{4} & \red{4} & -8\tikzmark{right} & 0 & 0 & 0 & 0 & 0 & 0 \\
 \blue{-1-71 i} & \blue{1+71 i} & 0 & -8-46 i & 0 & 0 & 0 & 0 & 0 \\
 0 & 0 & 0 & 0 & -9+12210 i & 0 & \blue{0+71 i} & 0 & 0 \\
 \blue{-1+71 i} & \blue{1-71 i} & 0 & 0 & 0 & -8+46 i & 0 & 0 & 0 \\
 0 & 0 & 0 & 0 & \blue{0+71 i} & 0 & -9+12256 i & 0 & 0 \\
 0 & 0 & 0 & 0 & 0 & 0 & 0 & -9-12210 i & \blue{0-71 i} \\
 0 & 0 & 0 & 0 & 0 & 0 & 0 & \blue{0-71 i} & -9-12256 i
\end{array}
\right)
\left( \begin{array}{c} \rho_{11} \\ \rho_{22} \\ \rho_{33} \\ \rho_{12} \\ \rho_{13} \\ \rho_{21} \\ \rho_{23} \\ \rho_{31} \\ \rho_{32} \end{array} \right)
\label{dimer superoperator with secular}
\end{align}
\DrawBox[thick]
\end{widetext}

The wide-spread usage of this full secular approximation has lead some to believe that the Bloch-Redfield formalism is not capable of modeling coherent oscillations at all. The partial secular approximation however leaves the non-negligible elements behind, which cause coherent oscillations. Reference \onlinecite{Ishizaki2009} discussed furthermore how the full secular approximation alters energy transfer rates between sites when the reorganisation energy becomes greater than the electronic couplings. References \onlinecite{Ishizaki2009, Shabani2012} criticise further that Bloch-Redfield equations in general are only second-order in the noise strength although the noise and the couplings are of similar order in light-harvesting systems. Note however, that this criticism applies equally to Lindblad equations as typically the Lindblad rates are derived from a second-order perturbation theory calculation of the system-environment coupling.

The secular approximation is particularly useful for finding analytical solutions. It should be emphasised that proper application of the secular approximation should result in the same dynamics as the original form of the Bloch-Redfield equations. For purely numerical simulations the Bloch-Redfield equations can therefore often be simulated without the need to apply the secular approximation. The approximation leaves the overall size of the superoperator unchanged and usually does not create computational speedup. The secular approximation is only relevant for numerical speedup if it is used to map to Lindblad equations with a subsequent mapping to a quantum jump algorithm. This can be useful for treating very large quantum systems\cite{Vogt2013}.

\section{Mapping to Lindblad equations}
\label{sec mapping to lindblad}
If a Lindblad form is preferred there are several ways by which one can map Bloch-Redfield equations to a Lindblad form, which guarantees by its mathematical form both positivity and the stronger property complete positivity, i.e.\ the eigenvalues of any larger matrix $\rho \otimes \mathds{1}$ also stay positive in the time evolution\cite{Breuerbook}. The correct approach for the mapping depends on the level of modeling detail that one wants to transmit to Lindblad form. We can map the Bloch-Redfield equations to Lindblad equations by neglecting all time correlations $C_{jk}(\omega)=C_{jk}$ (or taking the secular approximation) and neglecting all spatial correlations of the bath $C_{jk}\propto \delta_{jk}$ (or diagonalising the coefficient matrix). Neglecting all time correlations means assuming a strong form of Markovianity in that the bath correlations decay instantly and not just on a time scale shorter than the system dynamics. The spectral function must then be constant for all frequencies. This strong condition can however be replaced with the secular approximation combined with a piece-wise flat spectral function which only changes on the large scale of the secular approximation\cite{Vogt2013}. The system operators can then be split up $s_j=\sum_\epsilon s_j(\epsilon)$ into the parts which are dependent on the same flat piece of the spectral function $C_{jk}(\omega \approx \epsilon)$. The secular approximation then neglects all mixed terms of parts at different energies $\epsilon$ and the Bloch-Redfield equations \eqref{B-R equations} simplify:
\begin{align}
\dot \rho =& \frac{i}{\hbar} [\rho,H_s] \\&+ \frac{1}{\hbar^2} \sum_{j,k,\epsilon} \frac{1}{2} C_{jk}(\epsilon) \left(2 s_k(\epsilon) \rho s_j(\epsilon)^\dagger - \{ s_j(\epsilon)^\dagger s_k(\epsilon), \rho\} \right) \nonumber
\end{align}
This is discussed in more detail in reference \onlinecite{Vogt2013}, where we also show how to implement these equations in a computationally fast quantum jump approach.

Neglecting all spatial correlations in the noise sets all crossed terms involving $j\neq k$ to zero since the corresponding spectral functions $C_{jk}(\omega)=0$. For correlated environments a diagonalisation of the coefficient matrix $(C_{jk})$ is necessary\cite{Jeske2013formalism}. This however is a non-trivial step and will yield non-local Lindblad operators. The mapping is then merely a step to reassure physical behaviour of the master equation. The mapping shows that Bloch-Redfield equations are a very similar tool to Lindblad equations with the same range of validity, namely Markovian noise. On the other hand the mapping also highlights that the Lindblad form, obtained from a mapping, is often non-trivial and could not be easily formulated phenomenologically. The Bloch-Redfield formalism is therefore an excellent tool to model more complex environments by considering different types of correlations. 

Lindblad equations can be (and are often) derived from an underpinning model with similar techniques employed in the Bloch-Redfield formalism. However, if the Lindblad form is employed phenomenologically and purely for its mathematical properties the connection to a microscopic noise model can be lost. This can lead to artificial effects like noise-induced oscillations whose physical cause is unclear. These can be misinterpreted as system oscillations\cite{Palmieri2009} but a characteristic feature is that the strength of such oscillations is purely dependent on the noise strength and shows the same temperature dependency as the noise-induced decays.

\section{Effects in LHCs}
\label{sec effects in large LHCs}
\begin{figure}
%\flushleft\vspace{1.6cm}                            %\\
\includegraphics{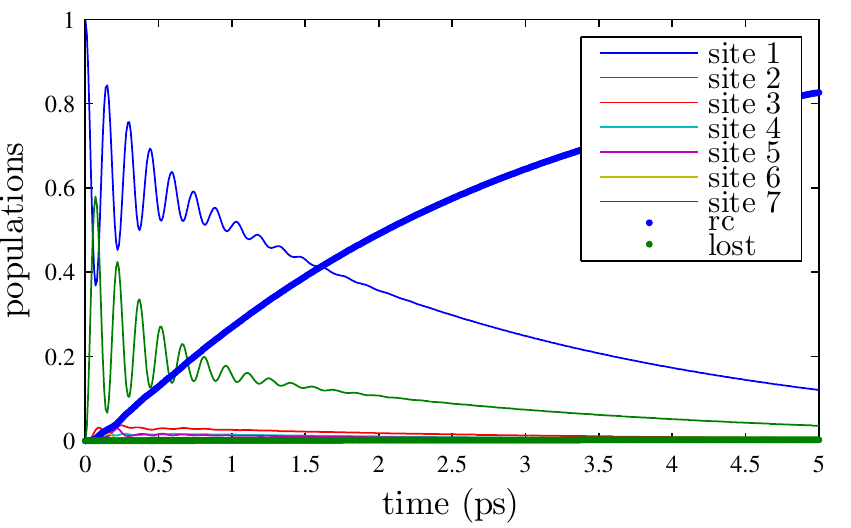}\\%\vspace{2.3cm}\\
\includegraphics{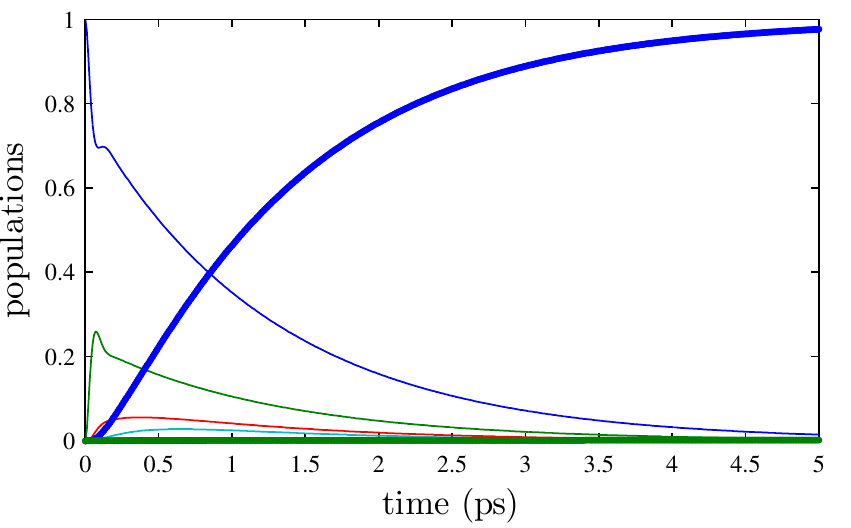}\\%\vspace{2.3cm}\\
\includegraphics{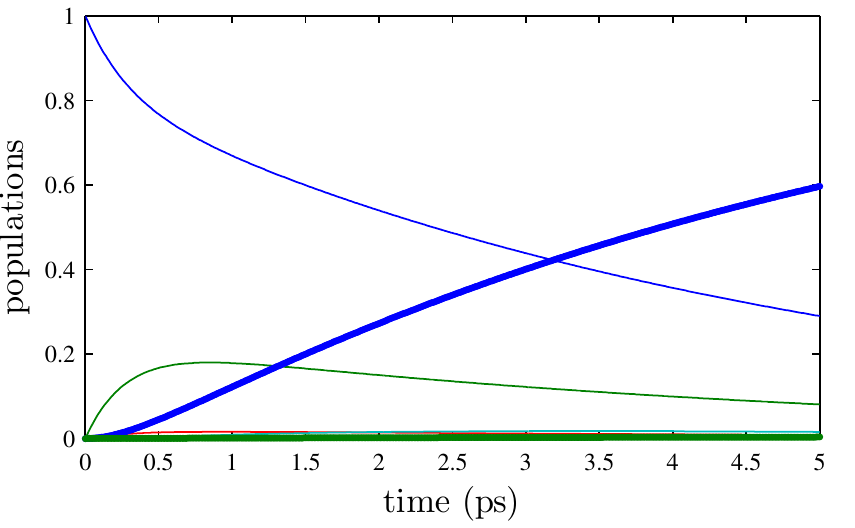}
\caption{Time evolution of the FMO complex at 77K and for three different dephasing rates: from top to bottom $\gamma_2=1$ps$^{-1},\,10$ps$^{-1},100$ps$^{-1}$. Top: Within the coherence time there are several oscillations between site 1 and 2. Middle: Around the optimal dephasing rate the excitation is transferred very quickly. Bottom: Strong dephasing starts to ``freeze'' the system (quantum zeno effect), which slows down the transfer. The same calculation without including a trapping site has been carried out in appendix \ref{app FMO without trapping}}
\label{fig FMO time traces}
\end{figure}

The Fenna-Matthews-Olson (FMO) complex is a seven site light-harvesting complex, in which the Hamiltonian governing the excitonic dynamics has been calculated \cite{Adolphs2006}, and which has therefore been studied by various groups\cite{Camara2003, Mohseni2008, Plenio2008, Thorwart2009, Rebentrost2009SpatialCor, Hoyer2010, Wu2010, Sarovar2011, Kolli2011, Ritschel2011}. In this section we demonstrate how the Bloch-Redfield equations provide an efficient tool to investigate LHCs such as the FMO complex, which is also supported by other studies, which used Bloch-Redfield equations in the context of LHCs \cite{Renger1998, Renger2002, Novoderezhkin2004, Kjellberg2006, Adolphs2006, Abramavicius2011, Rebentrost2009SpatialCor}. We employ the Hamiltonian for the FMO complex of Chlorobium tepidum as given in ref.~\onlinecite{Marais2013} and originally calculated in ref.~\onlinecite{Adolphs2006} and use site-numbering that follows the original paper by Fenna et al\cite{Fenna1975}. The system operators which couple to the noise and the spectral function are chosen analogously to the dimer in the previous section, with an additional trapping rate of $10$ps$^{-1}$ from site 3 to the reaction centre (rc). We simulate the FMO complex at $T=77$K and set the recombination rate to $\gamma_1=\nu^2 C(12210$cm$^{-1})=0.001$ps$^{-1}$. The time evolution for different dephasing rates $\gamma_2=v^2 C(0)$ is shown in figure \ref{fig FMO time traces}. The time scales compare to similar calculations by Adolphs et al\cite{Adolphs2006}, who also used a Bloch-Redfield approach to successfully reproduce experiments; however, contrary to that reference we plot in the site basis $\ket{j}$ to see the coherent spatial oscillations, rather than the exciton basis, i.e. the eigenbasis of the system Hamiltonian. 

We find that even for strong noise, compared to the system couplings, the solution shows physical behaviour, ie.\ non-positivity is not an issue. The occurrence of an optimal dephasing rate is in agreement with other models\cite{Plenio2008, Wu2010}. 

To demonstrate further the capabilities of the Bloch-Redfield equations we investigate the influence of temperature and spatial correlations on the transfer dynamics of the FMO. Recently Olbrich et al.\ did not find correlations within the system site energies in a classical molecular dynamics simulation of a truncated version of the FMO complex \cite{Olbrich2011}. This can be seen as an indication of uncorrelated noise,  however a full quantum mechanical model of the entire protein surrounding the FMO is beyond current computational capabilities. Furthermore Fokas et al.\ found highly correlated motions at very  low frequency between excitonically coupled elements more recently \cite{Fokas2014} employing constrained geometric dynamics.

\begin{figure}
\flushleft
%\vspace{1.6cm}                                %\\
\includegraphics{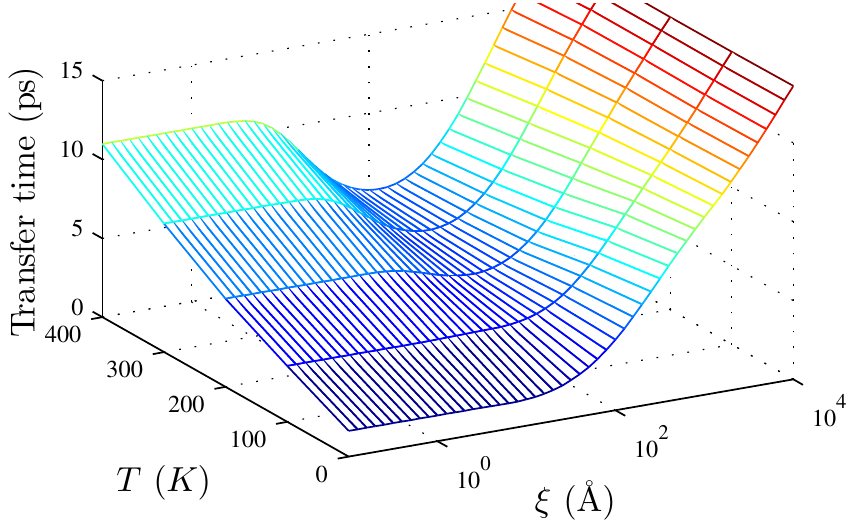}
\caption{Transfer time as a function of temperature and spatial noise correlation length for a dephasing rate $\gamma_2=10$ ps$^{-1}$. We find an optimal region with short transfer times for finite correlation length $\xi\approx100$ \AA. However this optimal value is very dependent on the dephasing rate, see figure \ref{fig transfer time dephasing vs xi}.}
\label{fig transfer time temp vs xi}
\end{figure}

Temperature is modelled via the Ohmic spectral function, figure~\ref{fig spectral}. The influence of spatial noise correlations is modelled, similar to refs.~\onlinecite{Adolphs2006, Rebentrost2009SpatialCor}, via a homogeneous exponentially decaying function (eq.~\ref{spatial correlations exp decay}), which we combine with the three-dimensional relative distances of the FMO chromophores\cite{Ben-Shem2004, FMOproteinDataBase}. Figure \ref{fig transfer time temp vs xi} shows the time for a 90\% probability of the excitation initially placed on site 1, to transfer to the reaction centre as a function of temperature and spatial correlation length $\xi$. For temperatures, which are not too close to zero we find that increasing the correlation length of the noise from zero enhances the excitation transport by reinstating the coherent transfer dynamics\cite{Jeske2013spinchain}. For $\gamma_2=20$ps$^{-1}$ the optimal correlation length is approximately $\xi=100$ \AA. Even longer correlation lengths are detrimental to the transfer. This is in agreement with the finding that a certain level of dephasing is advantageous to the transfer \cite{Plenio2008, Mohseni2008, Sinayskiy2012, Marais2013}. A correlation length of $100$~\AA \, may appear long compared to the site distances of the FMO complex which are between 10 and 30 \AA. However, since the correlations decay not as a step function but exponentially, the decay length $\xi=100$~\AA \,means that the noise between different pairs of sites show correlations between 90\% and 74\%. 

\begin{figure}
\flushleft
%\vspace{8cm}                                     %\\
\includegraphics[scale=0.055]{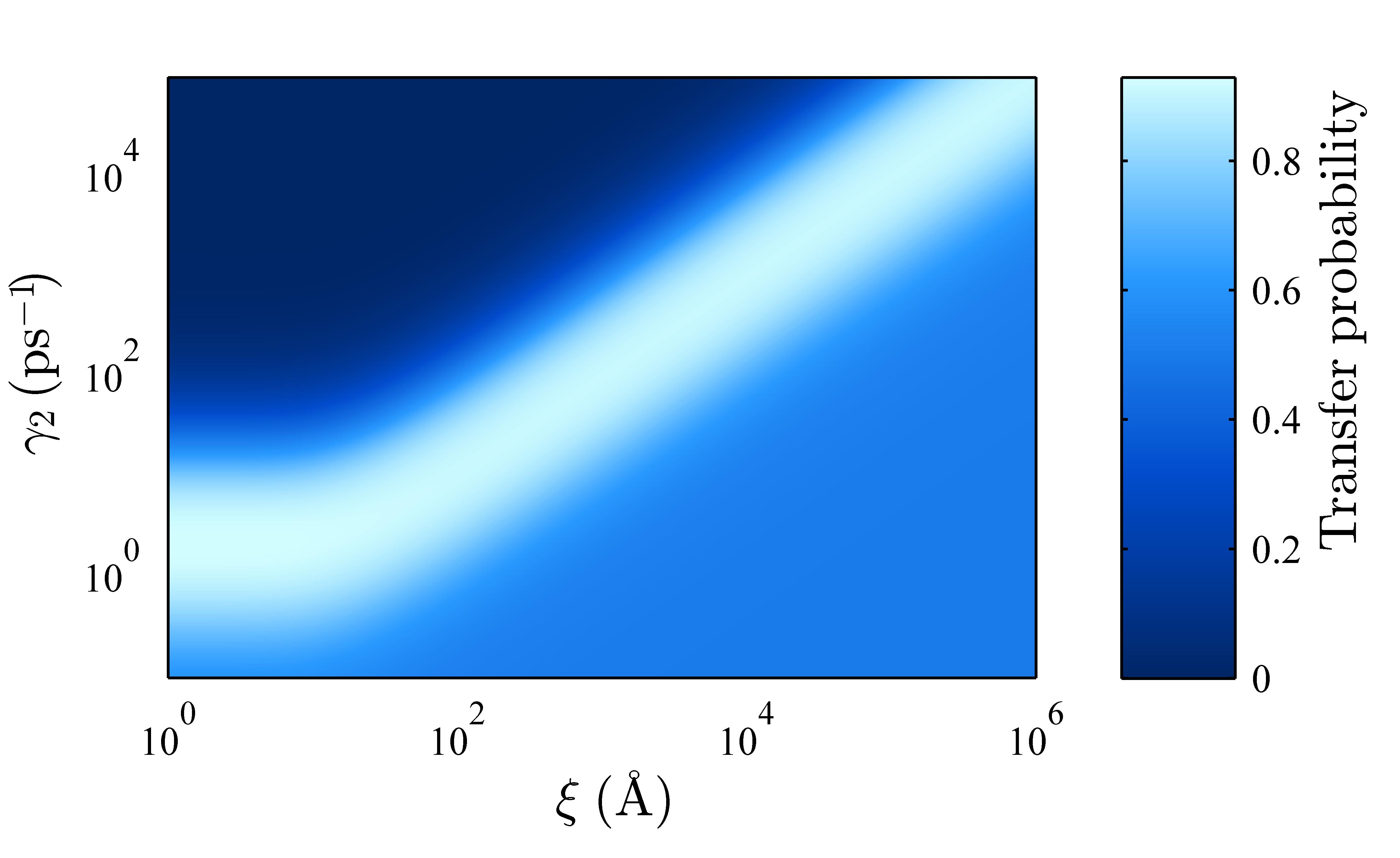}
\caption{Probability of excitation transfer to the reaction centre after $5$ ps$^{-1}$. Since correlation length reduces the dephasing noise effects, correlations can help to reach the optimal parameter regime and hence be advantageous to transport. Temperature was $T=277 K$ in this simulation.}
\label{fig transfer time dephasing vs xi}
\end{figure}

Increasing noise correlation length leads to dephasing-reduced subspaces of states with equal excitation number and transport processes are limited to the single-excitation subspace \cite{Jeske2013spinchain}. Therefore the optimal correlation length is strongly dependent on the dephasing rate. 
We find in figure \ref{fig transfer time temp vs xi} that around the optimal correlation length the transfer time can be reduced to less than $20$ps. We plot the probability of transfer after this time in figure \ref{fig transfer time dephasing vs xi} as a function of correlation length and dephasing rate. The dependence of the optimal correlation length on the dephasing rate is consistent over many orders of magnitude.

The Bloch-Redfield equations have proven to be a very useful tool for LHCs and to facilitate more comprehensive modeling options than phenomenological Lindblad equations. They provide further insight into the effects of environmental parameters, such as temperature, spatial correlations or the explicit form of the noise spectrum. The alleged weaknesses, namely, non-positivity, divergences or the inability to model coherent oscillations or noise which is as strong as the system couplings can all be overcome by a consistent underpinning physical model. We expect the Bloch-Redfield equations to be applicable and able to provide useful insight for several open questions in the field of LHCs and quantum biology in general, such as the role of vibrational modes in excitation transport, the detailed interpretation of 2D spectroscopy or the role of spatial noise correlations in biological transport dynamics.

\section{Conclusions}
We have presented how the Bloch-Redfield equations can be utilised to model excitation transfer dynamics in chromophoric aggregates or light-harvesting complexes in a consistent manner. The equations link back generally to a physical model of system-environment interaction Hamiltonian and spatial-temporal correlations contained in the spectral function, giving more flexible and adaptable modeling options than a phenomenological Lindblad approach. 

We have shown how issues of non-positivity and non-physicality or the loss of coherent oscillations are not inherent to the Bloch-Redfield formalism, and can be ruled out by an underpinning consistent physical model. If the secular approximation is applied, it needs to be based on the occurrence of different scales and applied carefully for the respective system at hand. Given these conditions, the secular approximation will not significantly alter the equations but merely simplify them by setting negligible elements to zero. 

We have illustrated the consistent use of the Bloch-Redfield formalism within two scenarios: a model dimer system and a prototypical LHC, the FMO complex. In the later 
we have combined a model with finite correlation length of the dephasing noise with the actual relative positions of the chromophores in the complex. We show the relative influences of correlation length, dephasing strength and temperature on the transfer time and probability. No issues of non-physicality arise, even for strong noise relative to the system couplings. We find an optimal noise correlation length, which is particularly relevant at higher temperatures, and strongly dependent on the dephasing rate. Our findings are both in agreement with and an extension of previous work by other groups. In conclusion the Bloch-Redfield equations provide an excellent tool to model Markovian noise in light-harvesting complexes when consistently applied, with the advantage of presenting an underlying microscopic model which can be viewed as a precursor for formulating higher order descriptions that include effects beyond the Markovian framework.

\acknowledgments
We acknowledge valuable discussions with N. Vogt. This work was supported by an Alexander von Humboldt Professorship, the EU STREP PAPETS and the EU Integrating Project SIQS.

%\clearpage
\appendix
%\section{Appendix}

\section{FMO without trapping}
\label{app FMO without trapping}
\begin{figure}
\flushleft %\vspace{1.8cm}                         %\\
\includegraphics{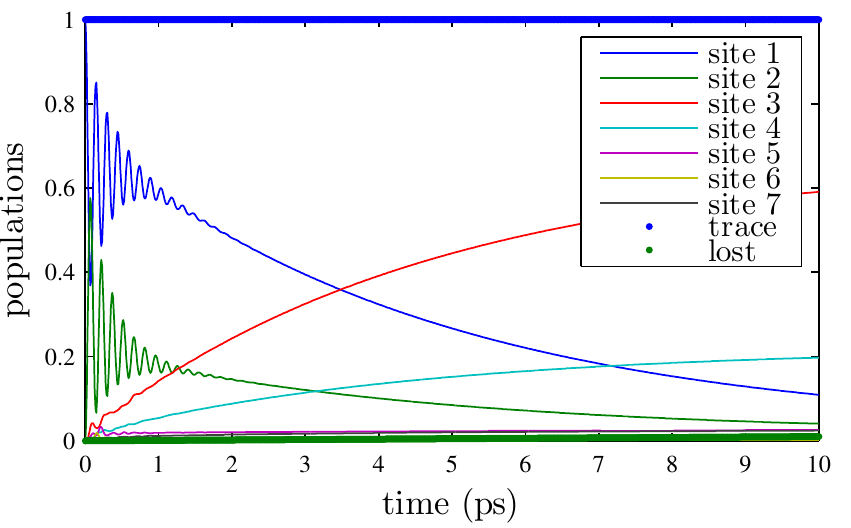}\\
\includegraphics{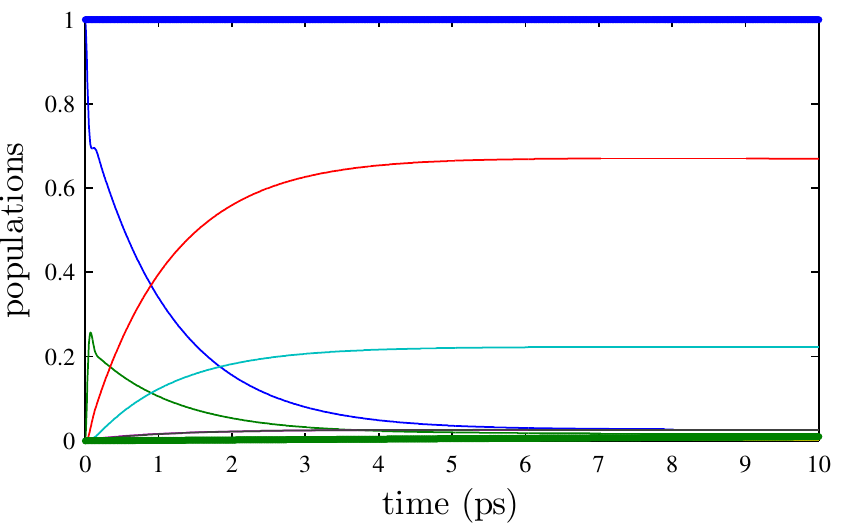}\\
\includegraphics{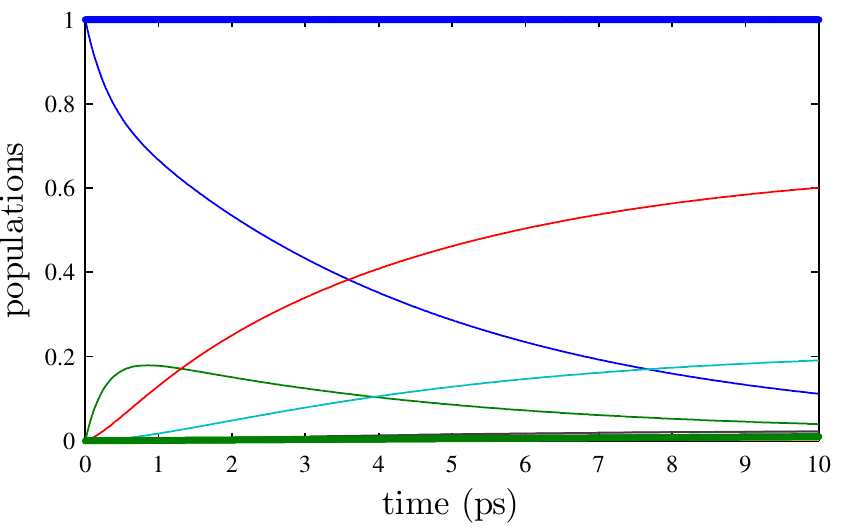}
\caption{Time evolution of the FMO complex without trapping to the reaction centre with otherwise the same parameters as fig.~\ref{fig FMO time traces}. The time evolution is trace-preserving. The oscillations are still damped by dephasing and the excitation ends up mainly on site 3 and 4. Other physical effects are analogous to fig.~\ref{fig FMO time traces}.}
\label{fig FMO time traces No Trapping}
\end{figure}

Throughout our FMO calculations we have assumed a trapping rate from site 3 to the reaction centre\cite{Rebentrost2009}, which distinguishes between two final states: the excitation has been transferred to the reaction centre or alternatively was lost through recombination. However trapping rates or absorption rates can in certain circumstances mask non-physical time evolution. To show that this is not the case we performed the calculations again without the trapping rate. Keeping otherwise the same parameters as in fig.~\ref{fig FMO time traces} we show the resulting time evolution in fig.~\ref{fig FMO time traces No Trapping} and plot additionally the trace of the density matrix $\sum_j \rho_{jj}=1$ which is preserved throughout the time evolution.

\begin{figure}
\flushleft                          %\\
\includegraphics{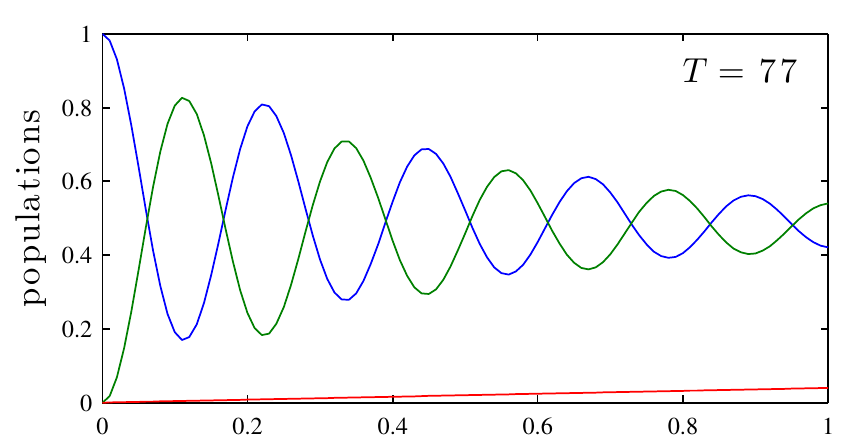}\\
\includegraphics{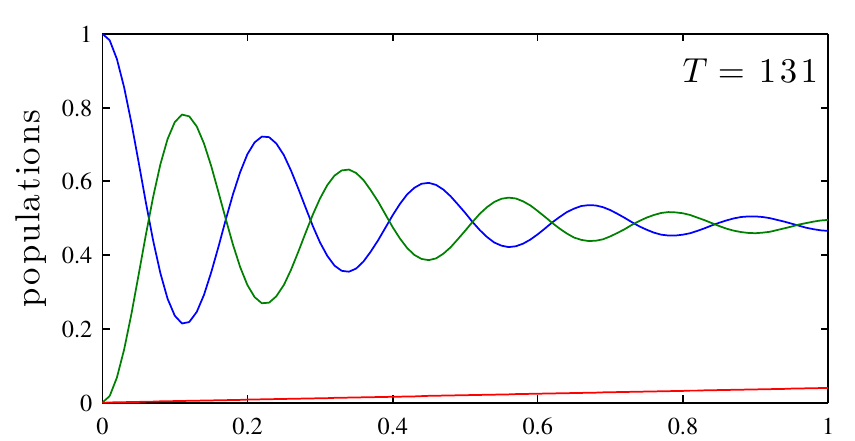}\\
\includegraphics{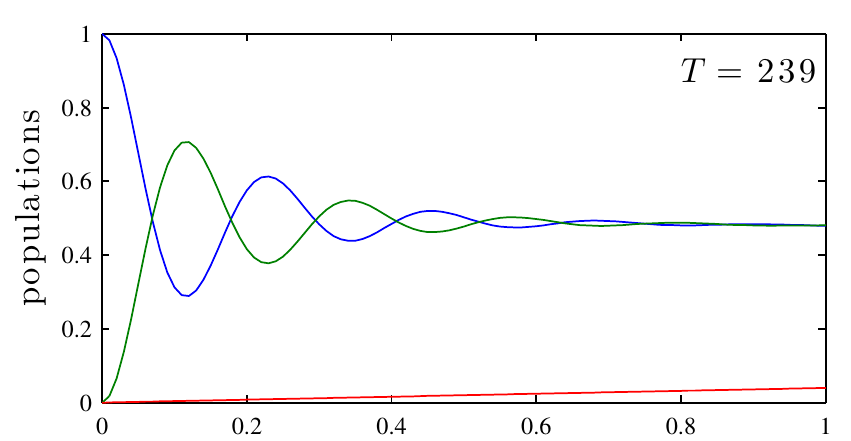}\\
\includegraphics{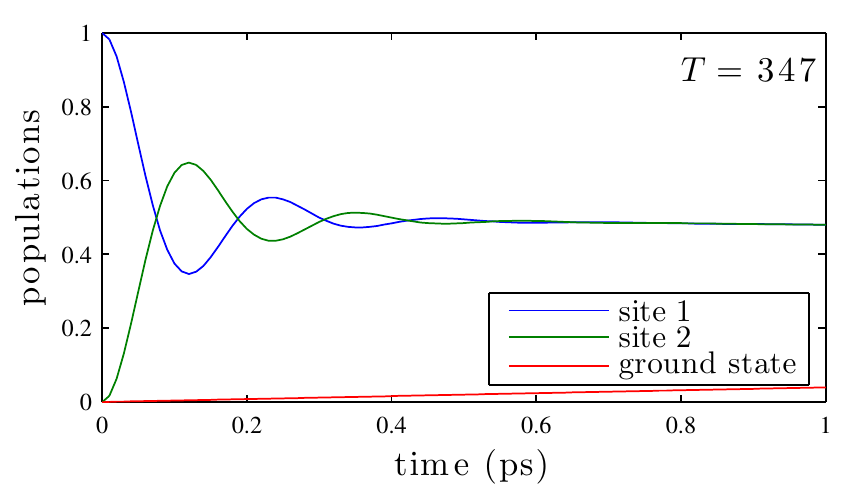}
\caption{Reproduction of figure \ref{fig dimer evolution} with phenomenological Lindblad equations. Since the spectral function is not used, the detailed balance is not enforced and the dephasing equalises populations in the sites.}
\label{fig Lindblad1}
\end{figure}

\begin{figure}
\flushleft	%\vspace{1.6cm}                         %\\
\includegraphics{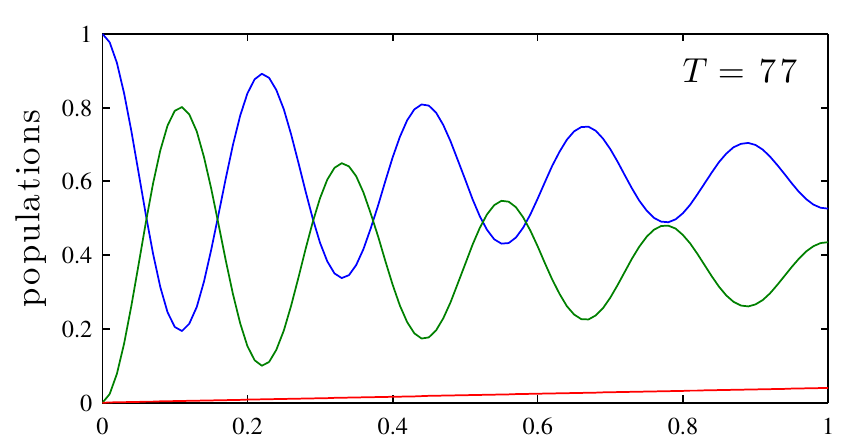}\\
\includegraphics{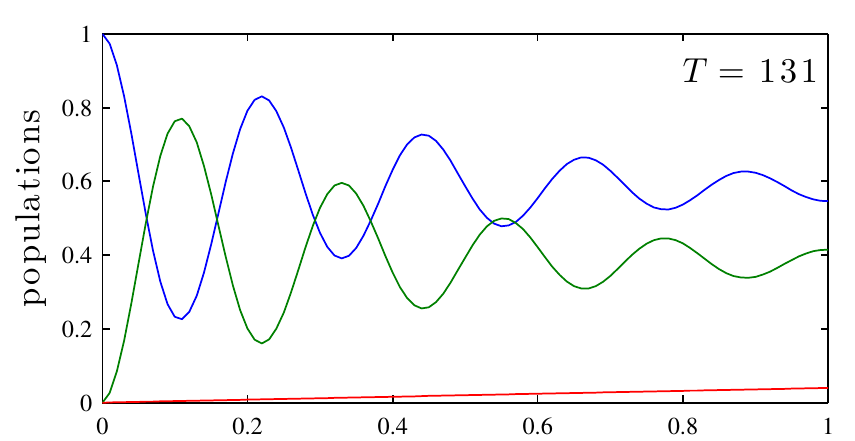}\\\includegraphics{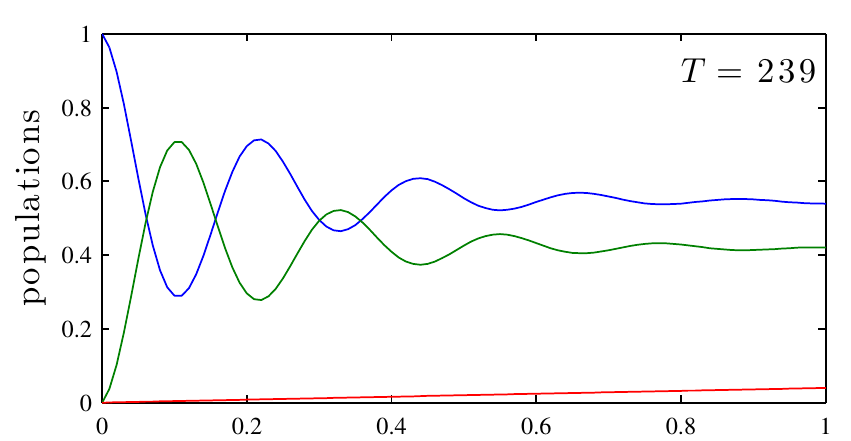}\\\includegraphics{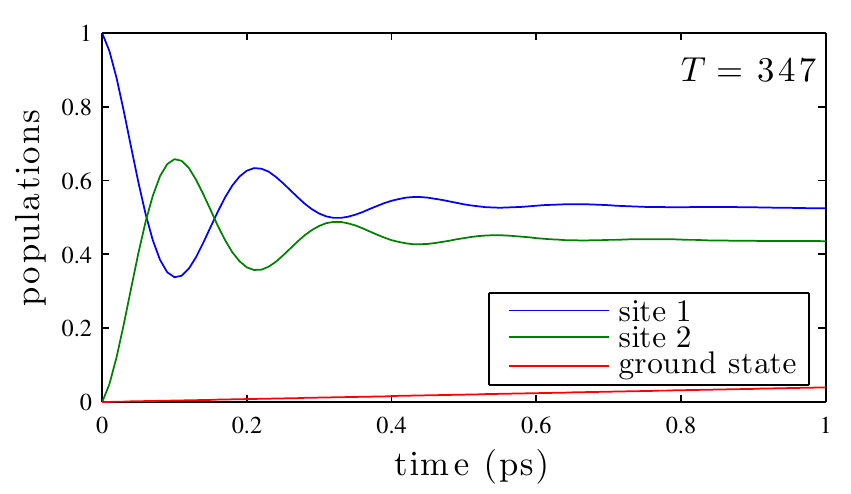}
\caption{A further reproduction of figure \ref{fig dimer evolution} with Lindblad equations as they arise from a formal mapping of the Bloch-Redfield equations. The differences to B-R are negligible. The Lindblad operators and rates arise from the mapping. }
\label{fig Lindblad2}
\end{figure}

\section{Reproduction of figure \ref{fig dimer evolution} with Lindblad equations}
We can reproduce figure \ref{fig dimer evolution} with Lindblad equations. We do so in two different ways. First we take phenomenological Lindblad equations. We assume Lindblad operators for dephasing on each site $L_1=2\ket{1}\bra{1}-\mathds{1}$ and $L_2=2\ket{2}\bra{2}-\mathds{1}$ and Lindblad operators for recombination on each site $L_3=\ket{1}\bra{0}$ and $L_4=\ket{2}\bra{0}$. The dephasing rate for both sites is equal and set to $\gamma_1=\gamma_2=v^2 C(\omega=0)$. This corresponds to the rate which occurs in B-R with the coupling strength $v$ and spectral function $C(\omega)$. Analogously the recombination rate is set to $\gamma_3=\gamma_4=\nu^2 C(\omega=12210\bf{cm}^{-1})$. The resulting evolution is displayed in figure \ref{fig Lindblad1}. It is very similar to the evolution under B-R, however the detailed balance is not enforced and the dephasing equalises the populations in both sides. This is because the spectral function is not applied in the detail of B-R. To achieve the detailed balance we also use Lindblad operators and rates as we obtain them from a detailed mapping to Lindblad form as outlined in section \ref{sec mapping to lindblad}. From this process we obtain more Lindblad operators, which are not phenomenological or intuitive any more. This new set of Lindblad operators leads to a different evolution shown in figure \ref{fig Lindblad2}. The differences to B-R are now negligible. 

We see that Lindblad equations and Bloch-Redfield are both valid approaches which yield the same results. The strength of the Bloch-Redfield equation compared to a purely phenomenological Lindblad approach is that a close connection to the underlying physics can be made and more comprehensive modeling options are easily accessible, such as preserving the detailed balance via the spectral function or modeling spatially correlated noise.

\bibliography{aaaPublication}
\end{document}